\documentclass[prl,aps,reprint,superscriptaddress,amsmath,amssymb]{revtex4-1}

\usepackage{graphicx}
\usepackage{dcolumn}
\usepackage{bm}
\usepackage{enumitem}
\usepackage{multirow}
\usepackage{color}

\begin{document}

\preprint{APS/123-QED}

\title{Dinucleon and Nucleon Decay to Two-Body Final States with no Hadrons in Super-Kamiokande}

\newcommand{\AFFicrr}{\affiliation{Kamioka Observatory, Institute for Cosmic Ray Research, University of Tokyo, Kamioka, Gifu 506-1205, Japan}}
\newcommand{\AFFkashiwa}{\affiliation{Research Center for Cosmic Neutrinos, Institute for Cosmic Ray Research, University of Tokyo, Kashiwa, Chiba 277-8582, Japan}}
\newcommand{\AFFipmu}{\affiliation{Kavli Institute for the Physics and
Mathematics of the Universe (WPI), The University of Tokyo Institutes for Advanced Study,
University of Tokyo, Kashiwa, Chiba 277-8583, Japan }}
\newcommand{\AFFmad}{\affiliation{Department of Theoretical Physics, University Autonoma Madrid, 28049 Madrid, Spain}}
\newcommand{\AFFubc}{\affiliation{Department of Physics and Astronomy, University of British Columbia, Vancouver, BC, V6T1Z4, Canada}}
\newcommand{\AFFbu}{\affiliation{Department of Physics, Boston University, Boston, MA 02215, USA}}
\newcommand{\AFFuci}{\affiliation{Department of Physics and Astronomy, University of California, Irvine, Irvine, CA 92697-4575, USA }}
\newcommand{\AFFcsu}{\affiliation{Department of Physics, California State University, Dominguez Hills, Carson, CA 90747, USA}}
\newcommand{\AFFcnm}{\affiliation{Department of Physics, Chonnam National University, Kwangju 500-757, Korea}}
\newcommand{\AFFduke}{\affiliation{Department of Physics, Duke University, Durham NC 27708, USA}}
\newcommand{\AFFfukuoka}{\affiliation{Junior College, Fukuoka Institute of Technology, Fukuoka, Fukuoka 811-0295, Japan}}
\newcommand{\AFFgifu}{\affiliation{Department of Physics, Gifu University, Gifu, Gifu 501-1193, Japan}}
\newcommand{\AFFgist}{\affiliation{GIST College, Gwangju Institute of Science and Technology, Gwangju 500-712, Korea}}
\newcommand{\AFFuh}{\affiliation{Department of Physics and Astronomy, University of Hawaii, Honolulu, HI 96822, USA}}
\newcommand{\AFFicl}{\affiliation{Department of Physics, Imperial College London , London, SW7 2AZ, United Kingdom }}
\newcommand{\AFFkek}{\affiliation{High Energy Accelerator Research Organization (KEK), Tsukuba, Ibaraki 305-0801, Japan }}
\newcommand{\AFFkobe}{\affiliation{Department of Physics, Kobe University, Kobe, Hyogo 657-8501, Japan}}
\newcommand{\AFFkyoto}{\affiliation{Department of Physics, Kyoto University, Kyoto, Kyoto 606-8502, Japan}}
\newcommand{\AFFliv}{\affiliation{Department of Physics, University of Liverpool, Liverpool, L69 7ZE, United Kingdom}}
\newcommand{\AFFmiyagi}{\affiliation{Department of Physics, Miyagi University of Education, Sendai, Miyagi 980-0845, Japan}}
\newcommand{\AFFnagoya}{\affiliation{Institute for Space-Earth Environmental Research, Nagoya University, Nagoya, Aichi 464-8602, Japan}}
\newcommand{\AFFkmi}{\affiliation{Kobayashi-Maskawa Institute for the Origin of Particles and the Universe, Nagoya University, Nagoya, Aichi 464-8602, Japan}}
\newcommand{\AFFpol}{\affiliation{National Centre For Nuclear Research, 00-681 Warsaw, Poland}}
\newcommand{\AFFsuny}{\affiliation{Department of Physics and Astronomy, State University of New York at Stony Brook, NY 11794-3800, USA}}
\newcommand{\AFFokayama}{\affiliation{Department of Physics, Okayama University, Okayama, Okayama 700-8530, Japan }}
\newcommand{\AFFosaka}{\affiliation{Department of Physics, Osaka University, Toyonaka, Osaka 560-0043, Japan}}
\newcommand{\AFFox}{\affiliation{Department of Physics, Oxford University, Oxford, OX1 3PU, United Kingdom}}
\newcommand{\AFFqmul}{\affiliation{School of Physics and Astronomy, Queen Mary University of London, London, E1 4NS, United Kingdom}}
\newcommand{\AFFregina}{\affiliation{Department of Physics, University of Regina, 3737 Wascana Parkway, Regina, SK, S4SOA2, Canada}}
\newcommand{\AFFseoul}{\affiliation{Department of Physics, Seoul National University, Seoul 151-742, Korea}}
\newcommand{\AFFsheff}{\affiliation{Department of Physics and Astronomy, University of Sheffield, S3 7RH, Sheffield, United Kingdom}}
\newcommand{\AFFshizuokasc}{\affiliation{Department of Informatics in
Social Welfare, Shizuoka University of Welfare, Yaizu, Shizuoka, 425-8611, Japan}}
\newcommand{\AFFstfc}{\affiliation{STFC, Rutherford Appleton Laboratory, Harwell Oxford, and Daresbury Laboratory, Warrington, OX11 0QX, United Kingdom}}
\newcommand{\AFFskk}{\affiliation{Department of Physics, Sungkyunkwan University, Suwon 440-746, Korea}}
\newcommand{\AFFtokyo}{\affiliation{The University of Tokyo, Bunkyo, Tokyo 113-0033, Japan }}
\newcommand{\AFFtodai}{\affiliation{Department of Physics, University of Tokyo, Bunkyo, Tokyo 113-0033, Japan }}
\newcommand{\AFFtit}{\affiliation{Department of Physics,Tokyo Institute of Technology, Meguro, Tokyo 152-8551, Japan }}
\newcommand{\AFFtus}{\affiliation{Department of Physics, Faculty of Science and Technology, Tokyo University of Science, Noda, Chiba 278-8510, Japan }}
\newcommand{\AFFtoronto}{\affiliation{Department of Physics, University of Toronto, ON, M5S 1A7, Canada }}
\newcommand{\AFFtriumf}{\affiliation{TRIUMF, 4004 Wesbrook Mall, Vancouver, BC, V6T2A3, Canada }}
\newcommand{\AFFtokai}{\affiliation{Department of Physics, Tokai University, Hiratsuka, Kanagawa 259-1292, Japan}}
\newcommand{\AFFtsinghua}{\affiliation{Department of Engineering Physics, Tsinghua University, Beijing, 100084, China}}
\newcommand{\AFFynu}{\affiliation{Faculty of Engineering, Yokohama National University, Yokohama, 240-8501, Japan}}
\newcommand{\AFFllr}{\affiliation{Ecole Polytechnique, IN2P3-CNRS, Laboratoire Leprince-Ringuet, F-91120 Palaiseau, France }}
\newcommand{\AFFbari}{\affiliation{ Dipartimento Interuniversitario di Fisica, INFN Sezione di Bari and Universit\`a e Politecnico di Bari, I-70125, Bari, Italy}}
\newcommand{\AFFnapoli}{\affiliation{Dipartimento di Fisica, INFN Sezione di Napoli and Universit\`a di Napoli, I-80126, Napoli, Italy}}
\newcommand{\AFFroma}{\affiliation{INFN Sezione di Roma and Universit\`a di Roma ``La Sapienza'', I-00185, Roma, Italy}}
\newcommand{\AFFpadova}{\affiliation{Dipartimento di Fisica, INFN Sezione di Padova and Universit\`a di Padova, I-35131, Padova, Italy}}

\AFFicrr
\AFFkashiwa
\AFFmad
\AFFbu
\AFFubc
\AFFuci
\AFFcsu
\AFFcnm
\AFFduke
\AFFllr
\AFFfukuoka
\AFFgifu
\AFFgist
\AFFuh
\AFFicl
\AFFbari
\AFFnapoli
\AFFpadova
\AFFroma
\AFFkek
\AFFkobe
\AFFkyoto
\AFFliv
\AFFmiyagi
\AFFnagoya
\AFFkmi
\AFFpol
\AFFsuny
\AFFokayama
\AFFosaka
\AFFox
\AFFqmul
\AFFregina
\AFFseoul
\AFFsheff
\AFFshizuokasc
\AFFstfc
\AFFskk
\AFFtokai
\AFFtokyo
\AFFtodai
\AFFipmu
\AFFtit
\AFFtus
\AFFtoronto
\AFFtriumf
\AFFtsinghua
\AFFynu

\author{S.~Sussman} \email[Corresponding author: ]{sarafs@bu.edu}
\AFFbu
\author{K.~Abe}
\AFFicrr
\AFFipmu
\author{C.~Bronner}
\AFFicrr
\author{Y.~Hayato}
\AFFicrr
\AFFipmu
\author{M.~Ikeda}
\AFFicrr
\author{K.~Iyogi}
\AFFicrr 
\author{J.~Kameda}
\AFFicrr
\AFFipmu 
\author{Y.~Kato}
\AFFicrr
\author{Y.~Kishimoto}
\AFFicrr
\AFFipmu 
\author{Ll.~Marti}
\AFFicrr
% ...


\collaboration{The Super-Kamiokande Collaboration}
\noaffiliation

\date{\today}

\begin{abstract}

Using 0.37 megaton$\cdot$years of exposure from the Super-Kamiokande detector, we search for 10 dinucleon and nucleon decay modes that have a two-body final state with no hadrons. These baryon and lepton number violating modes have the potential to probe theories of unification and baryogenesis. For five of these modes the searches are novel, and for the other five modes we improve the limits by more than one order of magnitude. No significant evidence for dinucleon or nucleon decay is observed and we set lower limits on the partial lifetime of oxygen nuclei and on the nucleon partial lifetime that are above $4\times 10^{33}$ years for oxygen via the dinucleon decay modes and up to about $4 \times 10^{34}$ years for nucleons via the single nucleon decay modes.
\end{abstract}
\pacs{Valid PACS appear here}

\maketitle


One of the biggest unanswered questions about our universe is the origin of the matter/antimatter asymmetry that we observe. Non-conservation of baryon number, $\mathcal{B}$, is one of the three necessary conditions to create a baryon asymmetry where none previously existed~\cite{Sakharov}. Since $\mathcal{B}$ is an accidental symmetry in the Standard Model (SM) of particle physics, observation of $\mathcal{B}$ violation would imply new physics beyond the Standard Model. Many theoretical extensions of the SM allow violation of $\mathcal{B}$ and/or lepton number, $\mathcal{L}$, and predict experimentally observable processes (see reviews in~\cite{2013snowmass} and \cite{2006Nath}). The searches for ten such $\mathcal{B}$-violating processes via nucleon or dinucleon decay in Super-Kamiokande are detailed in this Letter. Four of the eight dinucleon decay modes studied here have $\Delta(\mathcal{B-L})=-2$, with two nucleons decaying to a lepton and an antilepton. A scenario in which baryon asymmetry would remain after $\Delta(\mathcal{B-L})=-2$ decays in the early universe is discussed in Ref.~\cite{babu1}. Three of the eight dinucleon decay modes, with two nucleons decaying to two antileptons, violate each of $\mathcal{B}$ and $\mathcal{L}$ by two units, but conserve the quantity $(\mathcal{B-L})$. These modes are interesting in the context of models such as~\cite{1985Alberico, 1982Arnellos, 1982Mohapatra, 2009Ajaib}, and are shown in Ref.~\cite{2015Learned} to be competitive with LHC measurements in probing the mass scale of new physics. The final dinucleon decay mode and the two single-nucleon decay modes studied here are radiative; these decay modes can arise in various models of grand unification, but are often predicted to have suppressed decay rates~\cite{1982Lucha, 1983Lucha}. The radiative modes 
have similar experimental signatures as the other modes studied; they also have similar signatures to the previously searched $p \rightarrow e^{+}\pi^{0}$ and $p \rightarrow \mu^{+}\pi^{0}$ modes, but have the benefit of higher detection efficiency due to the lack of hadronic interactions.

\begin{figure}[htb]
\includegraphics[width=\columnwidth, trim={5cm 6cm 11cm 12.5cm}, clip]{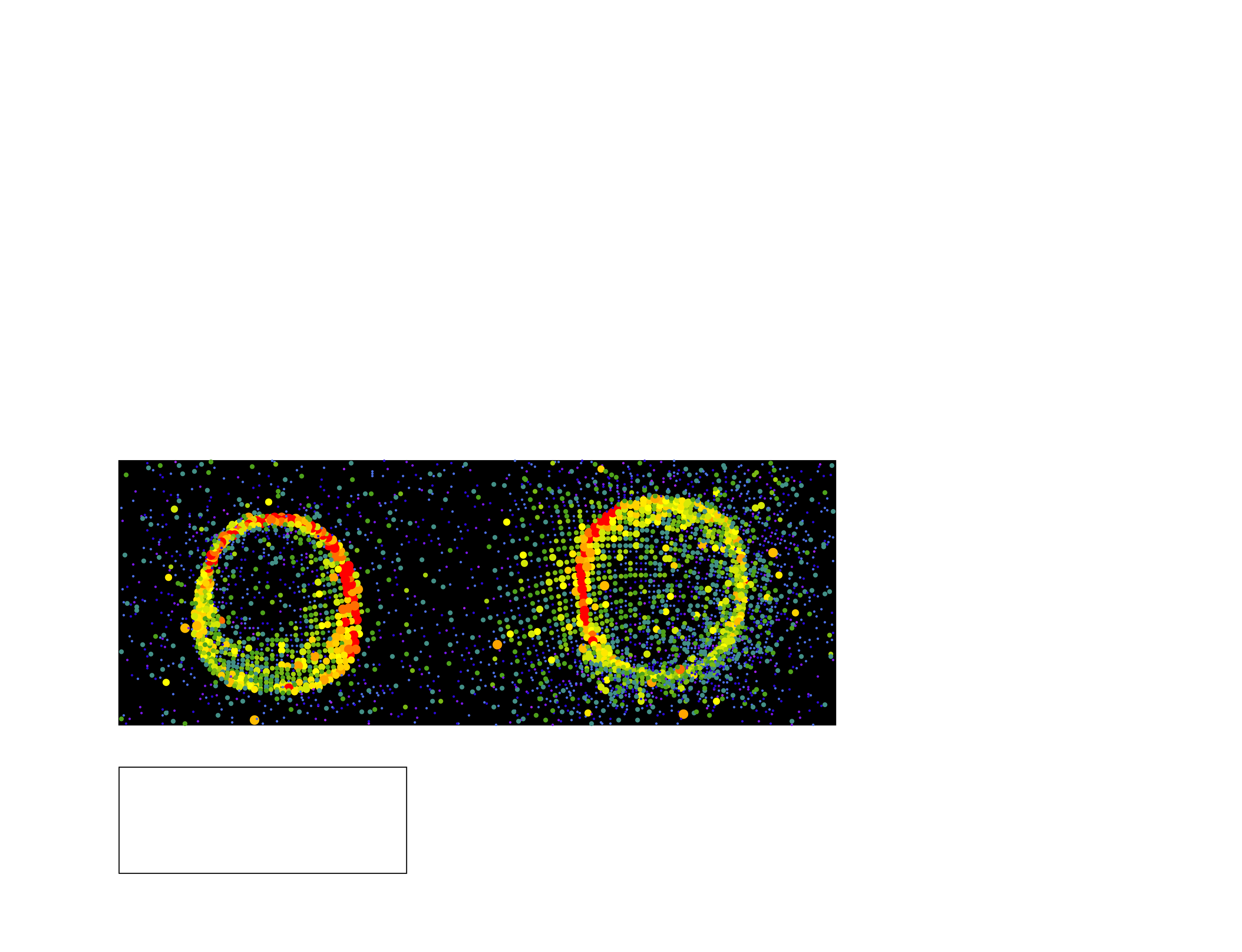}
\caption{(color online) An SK event display of a typical $pp\rightarrow e^{+}\mu^{+}$ event shown in $\theta$-$\phi$ view. The non-showering ring (from the $\mu^{+}$) is on the left and the showering ring (from the $e^{+}$) is on the right. The energy of each ring is approximately 900~MeV.}
\label{fig:evtdisp}
\end{figure}

The ten decay modes we search for in Super-Kamiokande data are characterized by two back-to-back Cherenkov rings and no hadrons. The dinucleon decay modes in these three categories are: (i) $pp \rightarrow e^{+}e^{+}$, $nn \rightarrow e^{+}e^{-}$, $nn \rightarrow \gamma\gamma$, (ii) $pp \rightarrow e^{+}\mu^{+}$, $nn \rightarrow e^{+}\mu^{-}$, $nn \rightarrow e^{-}\mu^{+}$, and (iii) $pp \rightarrow \mu^{+}\mu^{+}$, $nn \rightarrow \mu^{+}\mu^{-}$.  We classify the modes as follows: (i) both rings are showering ($NN\rightarrow ee$), (ii) one ring is showering and the other is non-showering ($NN\rightarrow e\mu$), and (iii) both rings are non-showering ($NN\rightarrow \mu \mu$). Figure~\ref{fig:evtdisp} illustrates how distinct these final states are seen in Super-Kamiokande, due to their well-separated, bright rings. The nucleon decay modes with identical experimental signatures, but lower invariant mass, are (i) $p \rightarrow e^{+}\gamma$ and (ii) $p \rightarrow \mu^{+}\gamma$. We do not include the search for dinucleon decays into tau leptons because there would be missing momentum and some subsequent tau decay modes are hadronic. 

The Super-Kamiokande (SK) water Cherenkov detector, with a fiducial volume of 22.5 kilotons, contains $1.2\times10^{34}$ nucleons. SK lies one kilometer under Mt. Ikenoyama in Japan's Kamioka Observatory. The detector is cylindrical with a diameter of 39.3 meters and a height of 41.4 meters, optically separated into an inner and an outer region. Eight-inch photomultiplier tubes (PMTs) line the outer detector facing outwards and serve primarily as a veto for cosmic ray muons, and 20-inch PMTs face inwards to measure Cherenkov light in the inner detector~\cite{SKNIM}. 

SK has collected data for four different detector periods, accumulating 91.5, 49.1, 31.8 and 199.3 kiloton$\cdot$years of exposure during SK-I, SK-II, SK-III, and SK-IV, respectively. During SK-I, the inner detector photocathode coverage was 40$\%$, but the SK-II period had a reduced photo-coverage of 19$\%$ after recovery from an accident. For SK-II, the remaining PMTs were evenly distributed to maintain isotropic detector uniformity. SK-II efficiency is only $\sim$2$\%$ lower than the other detector periods for these dinucleon and nucleon decay searches because the rings still have many hits. In the subsequent periods, SK-III and SK-IV, we restored the original photo-coverage of 40$\%$. The SK-IV period benefited from an electronics upgrade described in Ref.~\cite{2010elec}: a ``deadtime free'' data acquisition system enables SK-IV to detect the 2.2 MeV gamma ray emission from neutron capture on hydrogen, which occurs about 200~$\mu$sec after the primary event.

For each dinucleon or nucleon decay mode studied, we simulated 100,000 signal Monte Carlo (MC) events with vertices uniformly distributed throughout the detector and final state particle momenta uniformly distributed in phase space. Fermi motion, nuclear binding energy, and correlated decay are simulated in the dinucleon and nucleon decay signal MC~\cite{2017mine,2015jeff}. Unlike the atmospheric $\nu$ MC, where the Fermi momentum distribution of the nucleons follows the Fermi gas model, the signal MC Fermi momentum distribution follows a spectral function fit to electron-$^{12}$C scattering data \cite{197612c}. We address this difference between signal and atmospheric $\nu$ event samples by computing the systematic uncertainty in signal efficiency based on our choice of nuclear model. Correlated decay is a hypothesized effect where the total mass and momentum distributions are smeared out in a ``tail" due to the correlated motion of a nearby nucleon. For both nucleons and pairs of nucleons, we assume that 10$\%$ of such decays are affected by the correlated motion of an additional nucleon~\cite{1999corr}. Lepton rescattering within the nucleus is negligible. 

The atmospheric $\nu$ MC sample corresponds to an exposure of 500 years for each of the four SK periods, 2000 years in total. Events in this sample are weighted assuming two-flavor mixing as is done in recent dinucleon and nucleon analyses~\cite{2017mine,2016miura,2015jeff}. Details of the cross-sections and flux modeling used in this sample are discussed in recent SK nucleon decay analyses~\cite{2016miura, 2017mine}. Event rates obtained from this sample are normalized to the relevant SK detector livetime.

Details of the neutron simulation and neutron tagging algorithm used for both the signal and atmospheric $\nu$ MC samples can be found in Ref.~\cite{2016miura}. Neutron tagging can only be done for the SK-IV period; it reduces the expected number of background events by about 50\% for our searches, and impacts signal efficiency by only a few percent.

\begin{figure*}[htbp]
\centering
\includegraphics[scale=0.5, clip]{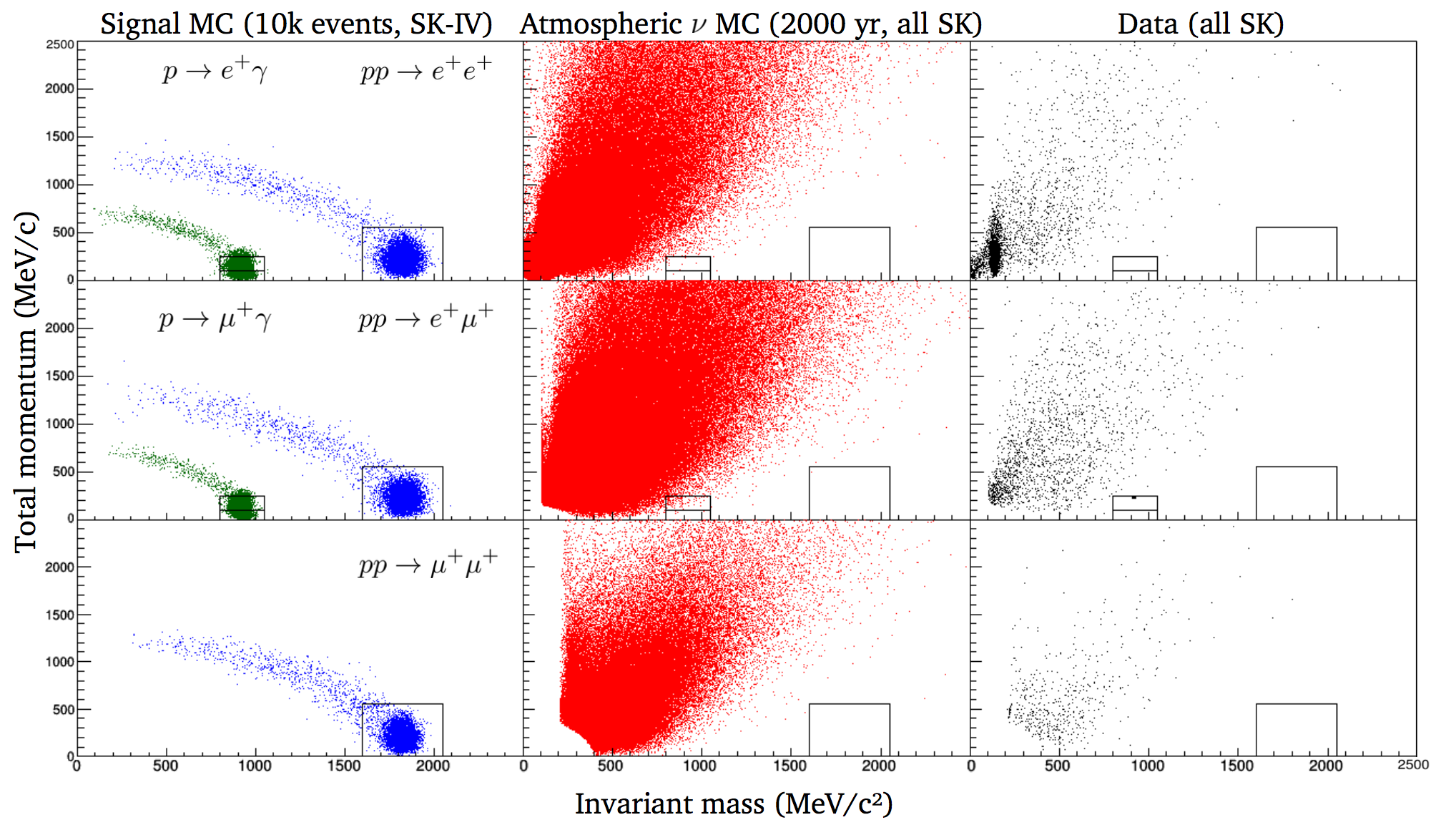}
\caption{(color online) Invariant mass vs. total momentum for several dinucleon and nucleon decay modes after cut (A4). The left panels show signal MC, where green corresponds to SK-IV nucleon decay MC and blue corresponds to SK-IV dinucleon decay MC for the labeled modes. The signal MC distributions for all SK periods look similar; only 10,000 signal MC events are plotted for each mode in order to more clearly show the shape of the distribution. The middle panels show atmospheric $\nu$ MC corresponding to 2000 years of SK exposure, and the right panels show SK-I through SK-IV data. The marker size is enlarged for data in the signal boxes.}
\label{fig:mass-momentum}
\end{figure*}

\begin{figure}[htbp!]
\centering
\includegraphics[scale=0.41, clip]{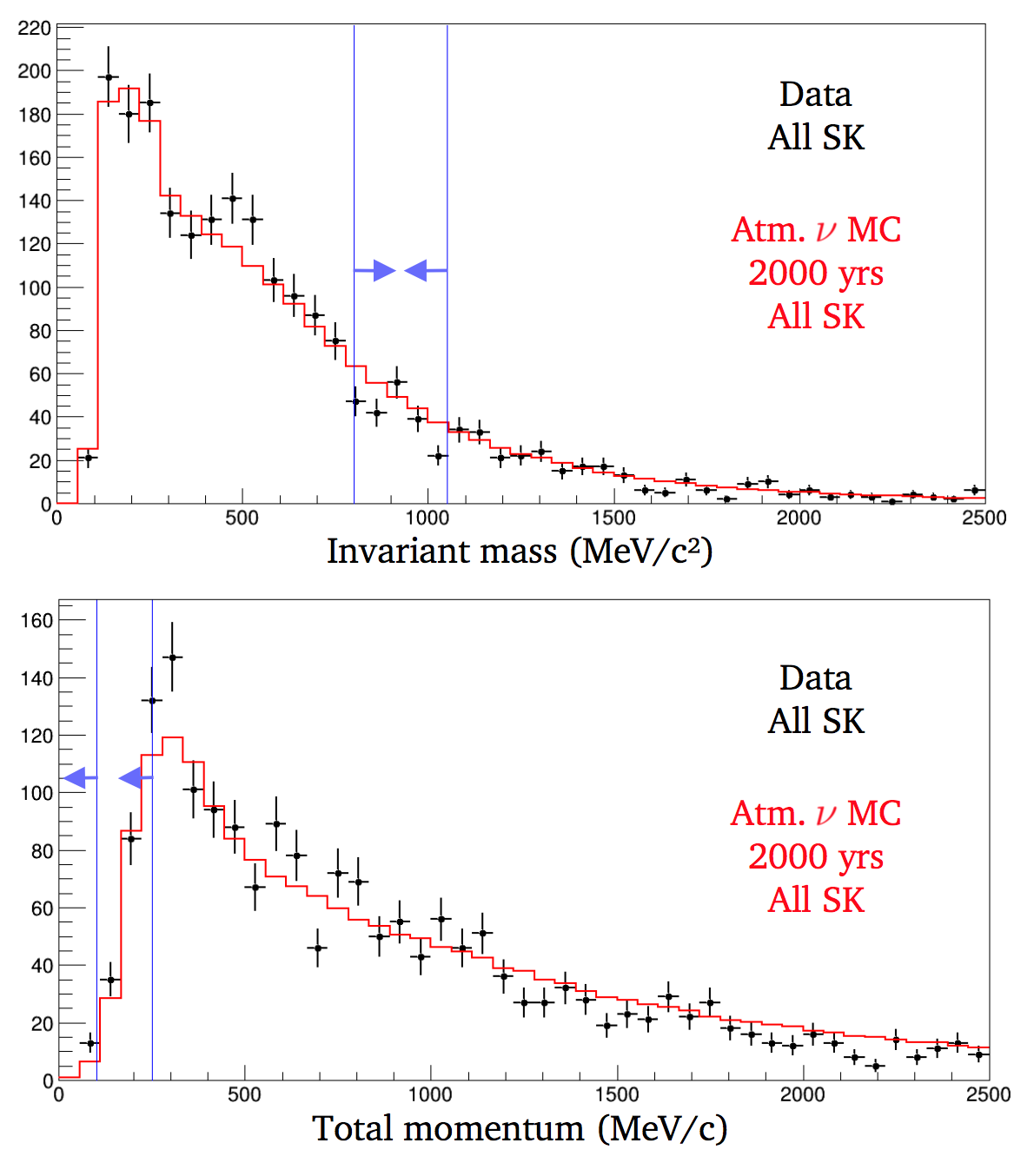}
\caption{(color online) Total mass ($M_{tot}$) and total momentum ($P_{tot}$) projections for $p \rightarrow \mu^+ \gamma$ after cut (A4). The red histogram shows atmospheric $\nu$ MC corresponding to 2000 years of SK exposure normalized to SK-I through SK-IV data. The selection criteria are indicated by the vertical blue lines.}
\label{fig:projection}
\end{figure}

Although the selection criteria for all ten modes are similar, the two single-nucleon decay modes have more background due to their lower total mass. We adapt our strategy, as is done in Ref.~\cite{2016miura}, to perform a two-box analysis which allows us to study free and bound protons separately.

The following selection criteria are applied to signal MC, atmospheric $\nu$ MC, and data:
\begin{enumerate}[label=(A\arabic*)]
\item{Events must be fully contained in the inner detector with the event vertex within the fiducial volume (two meters inward from the detector walls),}
\item{There must be two Cherenkov rings,}
\item{Both rings must be showering for the $pp \rightarrow e^{+}e^{+}$, $nn \rightarrow e^{+}e^{-}$, $nn \rightarrow \gamma\gamma$ and $p \rightarrow e^{+}\gamma$ modes; one ring must be showering and one ring must be non-showering for the $pp \rightarrow e^{+}\mu^{+}$, $nn \rightarrow e^{+}\mu^{-}$, $nn \rightarrow e^{-}\mu^{+}$ and $p \rightarrow \mu^{+}\gamma$ modes; both rings must be non-showering for the $pp \rightarrow \mu^{+}\mu^{+}$, $nn \rightarrow \mu^{+}\mu^{-}$ modes (see note in~\footnote{A note on cut (A3): Due to the two photons in the final state of $nn \rightarrow \gamma\gamma$, there is no visible energy in the first radiation-length of the two showers: this slightly impacts particle identification. We are about $1\%$ less efficient at particle identification for $nn \rightarrow \gamma\gamma$ than we are for similar dinucleon decay modes $pp \rightarrow e^{+}e^{+}$ and $nn \rightarrow e^{+}e^{-}$.}),}
\item{There must be zero Michel electrons for the $pp \rightarrow e^{+}e^{+}$, $nn \rightarrow e^{+}e^{-}$, $nn \rightarrow \gamma\gamma$ and $p \rightarrow e^{+}\gamma$ modes; there must be less than or equal to one Michel electron for the $pp \rightarrow e^{+}\mu^{+}$, $nn \rightarrow e^{+}\mu^{-}$, $nn \rightarrow e^{-}\mu^{+}$ and $p \rightarrow \mu^{+}\gamma$ modes; there is no Michel electron cut for the  $pp \rightarrow \mu^{+}\mu^{+}$, $nn \rightarrow \mu^{+}\mu^{-}$ modes (see note in~\footnote{The decay electron cut is loose for the non-showering dinucleon decay modes because there is almost zero background to eliminate.}),}
\item{The reconstructed total mass, $M_{tot}$, should be $1600\leq M_{tot}\leq 2050$ ~MeV/c$^2$ for the dinucleon decay modes; the reconstructed total mass should be $800\leq M_{tot}\leq 1050$~MeV/c$^2$ for the nucleon decay modes,}
\item{The reconstructed total momentum, $P_{tot}$, should be $0 \leq P_{tot} \leq 550$~MeV/c for the dinucleon decay modes; for the nucleon decay modes, it should be $100 \leq P_{tot} \leq 250$~MeV/c for the event to be in the ``High $P_{\text{tot}}$" signal box and $0 \leq P_{tot} \leq 100$~MeV/c for the event to be in the ``Low $P_{\text{tot}}$" signal box,}
\item{[SK-IV nucleon decay searches only] There must be zero tagged neutrons.}
\end{enumerate}

Figure~\ref{fig:mass-momentum} shows the distributions of signal MC events (left panels), atmospheric neutrino background (middle), and data (right) as a function of $P_{tot}$ versus $M_{tot}$ after cut (A4). The signal selection efficiencies and background rates are summarized in Table~\ref{tab:effic-bg} for each of the decay modes and each of the SK running periods. The signal efficiency for the two nucleon decay modes is $\sim 50\%$ for the ``High $P_{\text{tot}}$"signal box and $\sim 28\%$ for the ``Low $P_{\text{tot}}$" signal box for each SK period. It is worth noting that these signal efficiencies are significantly higher than those of the similar event signature in the $p \rightarrow \ell^+ \pi^0$ decay mode searches. These differences are due to the fact that the $\pi^0$ undergoes nuclear effects before exiting the nucleus while the $\gamma$ does not. For the eight dinucleon decay modes, the signal efficiency is $\sim$$80\%$ for each SK period. Due to the high total mass required in (A5), the modes are virtually background-free (as shown in the middle panels of Fig.~\ref{fig:mass-momentum}). 

Background estimates are done in one of the two following ways, depending on the number of background events that fall in the signal box: (1) for signal regions that contain more than 10 events from 2000 years of atmospheric $\nu$ MC, the background is estimated by the traditional method of counting the number of events that fall inside the signal region; or, (2) for signal regions that are nearly background-free, an extrapolation method is used to estimate the expected background using the distribution of events nearby (but outside) the signal region. The background extrapolation is done by measuring the distance from the center of the signal box to the location of each nearby event in mass-momentum parameter space, and then fitting an exponential to the distribution of distances. Integration of the exponential function up to the radius which approximates the signal box (250 units in mass-momentum parameter space) gives the estimated background inside the signal region. A similar estimation method was done in Ref.~\cite{shiozawathesis}. Background rate for $p \rightarrow e^+ \gamma$ ``Low $P_{\text{tot}}$" is estimated by extrapolation to be 0.089 events/Mton$\cdot$yr; we take double this value (0.18 $\pm$ 0.18  events/Mton$\cdot$yr) as a conservative estimate of the background rate for this decay mode. Similarly, we extrapolate for all of the dinucleon decay modes, finding background rates of 0.008 ($NN \rightarrow ee$), 0.033 ($NN \rightarrow e \mu$), and 0.006 ($NN \rightarrow \mu \mu$) events/Mton$\cdot$yr. We conservatively take the largest of these and double it as our estimate of expected background for all of the dinucleon decay modes: $0.07\pm 0.07$ events/Mton$\cdot$yr.


\begin{table*}[htbp]
\centering
\begin{tabular}{l l @{\hspace{5mm}} cccc @{\hspace{5mm}}cccc}
 \hline \hline
 \noalign{\vskip 1mm}
\multicolumn{2}{c}{Decay mode} & \multicolumn{4}{c}{Efficiency (\%)} & \multicolumn{4}{c}{Background (Events/livetime)} \\
   & & SK-I & SK-II & SK-III & SK-IV & SK-I & SK-II & SK-III & SK-IV \\
   \hline
   \noalign{\vskip 1mm}
\multirow{2}{*}{$p \rightarrow e^+ \gamma$} & High $P_{\text{tot}}$ & $51.0\pm0.2$  & $49.5 \pm 0.2$ & $50.8 \pm 0.2$ & $50.6 \pm 0.2$ & $0.01 \pm 0.01$ & $0.02 \pm 0.02$ & $< 0.01$ & $0.07 \pm 0.07$ \\
& Low $P_{\text{tot}}$ & $27.6 \pm 0.1$ & $26.1 \pm 0.1$ & $27.6 \pm 0.1$ & $27.5 \pm 0.1$ &  $0.02 \pm 0.02 $ & $0.01 \pm 0.01$ &  $0.01 \pm 0.01$ & $0.04 \pm 0.04$ \\
\noalign{\vskip 2mm}
\multirow{2}{*}{$p \rightarrow \mu^+ \gamma$} & High $P_{\text{tot}}$ & $50.2\pm 0.2$ & $49.7\pm0.2$ & $51.0 \pm 0.2$ & $48.1 \pm 0.2$ & $0.22 \pm 0.14$ & $0.14 \pm 0.11$ & $0.07 \pm 0.07$ & $0.23 \pm 0.14$\\
& Low $P_{\text{tot}}$ & $29.1 \pm 0.1$ & $28.3 \pm 0.1$ & $29.0 \pm 0.1$ & $29.4 \pm 0.1$ & $0.02 \pm 0.02$ & $0.01 \pm 0.01$ & $<0.01$  & $0.02 \pm 0.02$ \\
\noalign{\vskip 2mm}
$NN \rightarrow ee$ & & $80.9 \pm 0.1$ & $77.2 \pm 0.1$ & $79.5 \pm 0.1$ & $78.6 \pm 0.1$ & $0.01 \pm 0.01$ & $<0.01$ & $<0.01$ & $0.01 \pm 0.01$ \\ 
\noalign{\vskip 2mm}
$NN \rightarrow e\mu$ & & $84.1 \pm 0.1$ & $83.7 \pm 0.1$ & $83.4 \pm 0.1$ & $81.7 \pm 0.1$ & $0.01 \pm 0.01$ & $<0.01$ & $<0.01$ & $0.01 \pm 0.01$  \\
\noalign{\vskip 2mm}
$NN \rightarrow \mu\mu$ & & $86.3 \pm 0.1$ & $85.9 \pm 0.1$ & $86.0 \pm 0.1$ & $82.8 \pm 0.1$ & $0.01 \pm 0.01$ & $<0.01$ & $<0.01$ & $0.01 \pm 0.01$  \\
\noalign{\vskip 1mm}
\hline 
\hline
\end{tabular}
\caption{\label{tab:effic-bg} Summary of the number of expected background events (with statistical uncertainty only) for the livetimes of SK-I (91.5 kiloton$\cdot$years), SK-II (49.1 kiloton$\cdot$years), SK-III (31.8 kiloton$\cdot$years), and SK-IV (199.3 kiloton$\cdot$years). The dinucleon decay efficiency/background rate for a group of modes is the average of the efficiencies/background rates for the individual modes (the efficiencies are similar in the same group of modes.)}
\end{table*}

We find zero candidate events for the eight dinucleon decay modes. For the nucleon decay mode $p \rightarrow e^+ \gamma$, we also find zero candidate events. We observe two candidate events during the SK-IV period for the $p \rightarrow \mu^+ \gamma$ decay mode in the ``High $P_{tot}$" signal box when $0.23\pm0.14_{stat}\pm0.07_{sys}$ events were expected. The Poisson probability to see two or more events in the SK-IV livetime given an expected rate of 0.23 events is 2.3\%. One of the two candidates was previously found in Ref.~\cite{2016miura}. The other candidate is more ambiguous since it lacks a Michel electron. This may be an indication that the event is due to a $\nu_{e}n \rightarrow e^{-}p$ charged-current quasielastic interaction, where the non-showering ring is due to a proton rather than a muon. Requiring a Michel electron would have eliminated this event, however such a requirement was not applied for the $p \rightarrow \mu^{+}\gamma$ mode in order to be consistent with cut (A4) for the dinucleon decay mode. Fig.~\ref{fig:projection} shows the agreement of data and atmospheric $\nu$ MC for $p \rightarrow \mu^+ \gamma$.

\begin{table*}[htbp]
\centering
\begin{tabular}{l l @{\hspace{5mm}} ccc @{\hspace{5mm}} c }
\hline \hline
\noalign{\vskip 1mm}
\multicolumn{2}{c}{Decay mode} & \multicolumn{3}{c}{Signal efficiency uncertainty (\%)} & Background rate uncertainty(\%) \\
\multirow{2}{*}{} & \multirow{2}{*}{} & \multirow{2}{*}{Reconstruction} & Correlated & Nuclear & \\
& & & Decay & Model & \\
   \hline
   \noalign{\vskip 1mm}
\multirow{2}{*}{$p \rightarrow e^+ \gamma$} & High $P_{\text{tot}}$ & 10.5 & 3.5 & 2.4 & 40.4 \\
& Low $P_{\text{tot}}$ & 8.1 & 2.9 & 5.3 & 100 \\
\noalign{\vskip 2mm}
\multirow{2}{*}{$p \rightarrow \mu^+ \gamma$} & High $P_{\text{tot}}$ & 10.3 & 3.5 & 3.7 & 31.0 \\
& Low $P_{\text{tot}}$ & 8.0 & 3.1 & 5.8 & 44.0 \\
\noalign{\vskip 2mm}
$NN \rightarrow ee$ & & 5.7 & 8.0 & --- & 100 \\
\noalign{\vskip 2mm}
$NN \rightarrow e\mu$ & & 2.6 & 8.4 & ---  & 100 \\
\noalign{\vskip 2mm}
$NN \rightarrow \mu\mu$ & & 4.4 & 8.7 & ---  & 100 \\
\noalign{\vskip 1mm}
\hline
\hline
\end{tabular}
\caption{\label{tab:systematics} Summary of the systematic uncertainties (percentage contribution) on signal efficiency and background rate for the nucleon and dinucleon decay searches. The uncertainties from SK-I to SK-IV are averaged by the live time.}
\end{table*}

\begin{table*}[htbp!]
\begin{equation}
\label{eq:probability}
P(\Gamma | n_i)\! = \! \int^{\lambda}\! \int^{\epsilon}\! \int^{b} \! \frac{e^{-(\Gamma\lambda_i(\lambda)\epsilon_i(\epsilon)+b_i(b))}(\Gamma \lambda_i(\lambda)\epsilon_i(\epsilon)+b_i(b))^{n_i}}{n_i !}\! 
P(\Gamma)\! P(\lambda_i(\lambda) | \lambda_{i,0}, \sigma_{\lambda_{i,0}})\!P(\epsilon_i(\epsilon) | \epsilon_{i,0}, \sigma_{\epsilon_{i,0}})\!P(b_i(b) | b_{i,0}, \sigma_{b_{i,0}})d\lambda\,d\epsilon\,db
\end{equation}
\end{table*}

Table~\ref{tab:systematics} summarizes the systematic uncertainties in the signal efficiency and in the background rate for each of the nucleon and dinucleon decay modes. The dominant contributions to uncertainty in the signal efficiency arise from uncertainties in the areas of reconstruction, correlated decay, and nuclear model. To assess the impact of differences in the reconstruction of data and MC, for every variable used in the selection,
we compute the percent shift of the atmospheric $\nu$ MC distribution necessary to minimize its chi-square against the corresponding data distribution. 
The cut value in the event selection is then shifted by that percentage and applied to the signal MC to recalculate the efficiency. The total systematic uncertainty due to reconstruction is calculated by
summing in quadrature the independent percent changes in signal efficiency due to each percent-shifted cut. For nucleon decay in the SK-IV period only, we also add in quadrature with the other reconstruction uncertainties an additional 10\% uncertainty due to neutron tagging, as was done in Ref.~\cite{2016miura}. This is the reason that the reconstruction uncertainties for nucleon decay are $\sim$6\% larger than the corresponding uncertainties for dinucleon decay. To estimate the uncertainty in the signal efficiency arising from uncertainties in correlated decay, we assume 100\% uncertainty on the correlated decay effect, reweight the correlated decay events accordingly, and recalculate the signal efficiency, taking the overall change in signal efficiency as the systematic uncertainty. The nuclear model uncertainty is estimated as the percent change in signal efficiency when the Fermi gas model is used to compute the true momentum of the protons within the signal MC events instead of the spectral function fit to data described earlier. 

The systematic uncertainty on the rate of background events is conservatively taken to be 100\% for decay modes where the background events are scarce (all dinucleon decay modes, and the $p \rightarrow e^+ \gamma$ ``Low $P_{\text{tot}}$" nucleon decay). For the other nucleon decay signal regions, we use an event-by-event database with uncertainty weights from 73 sources of background systematic uncertainty including uncertainties in flux, cross section and energy calibration, as described in the 2018 SK oscillation analysis~\cite{2018osc}.

Lifetime limits are computed using a Bayesian method, assuming that the SK-I through SK-IV datasets have correlated systematic uncertainties~\cite{2018pdg}. For the nucleon decay modes, the systematic uncertainties of the ``High $P_{\text{tot}}$" and ``Low $P_{\text{tot}}$" search boxes are treated as independent datasets with fully correlated systematic uncertainties. The conditional probability distribution for the decay rate is given by Eq.~\ref{eq:probability}, where $\Gamma$ is the decay rate and for dataset $i$, $\lambda_i$ is the exposure (given in proton-years for nucleon decay and in oxygen-years for dinucleon decay), $\epsilon_i$ is the efficiency, $b_i$ is the number of background events, and $n_i$ is the number of candidate events. Since the systematic errors are correlated for all datasets, integrating the prior probability distribution up to $b$ in some dataset implies that we integrate the prior distribution in dataset $i$ up to $b_i(b)$. 

We assume a Gaussian prior distribution $P(\lambda_i(\lambda) | \lambda_{i,0}, \sigma_{\lambda_{i,0}})$ for $\lambda_i$ with a mean value of $\lambda_{i,0}$ and $\sigma_{\lambda_{i,0}}$ given by the $1\%$ percent systematic uncertainty in exposure. We also assume Gaussian priors $P(\epsilon_i(\epsilon) | \epsilon_{i,0}, \sigma_{\epsilon_{i,0}})$ and $P(b_i(b) | b_{i,0}, \sigma_{b_{i,0}})$ for $\epsilon_i$ and $b_i$ with standard deviations set to the total percent systematic uncertainties in efficiency and background, respectively. To require a positive lifetime, $P(\Gamma)$ is 1 for $\Gamma \geq 0$ and otherwise 0. We calculate the upper bound of the decay rate $\Gamma_{\text{limit}}$ as in Eq.~\ref{eq:conflevel}, with a 90\% confidence level (CL):

\begin{table}[!htbp]
\begin{equation}
\label{eq:conflevel}
\text{CL} = \frac{\int_{\Gamma=0}^{\Gamma_{\text{limit}}}\prod_{i=1}^{N} P(\Gamma|n_i)d\Gamma}{\int_{\Gamma=0}^{\infty}\prod_{i=1}^{N} P(\Gamma|n_i)d\Gamma}.
\end{equation}
\end{table}

\begin{figure}[hbtp]
\centering
\includegraphics[scale=0.23, clip]{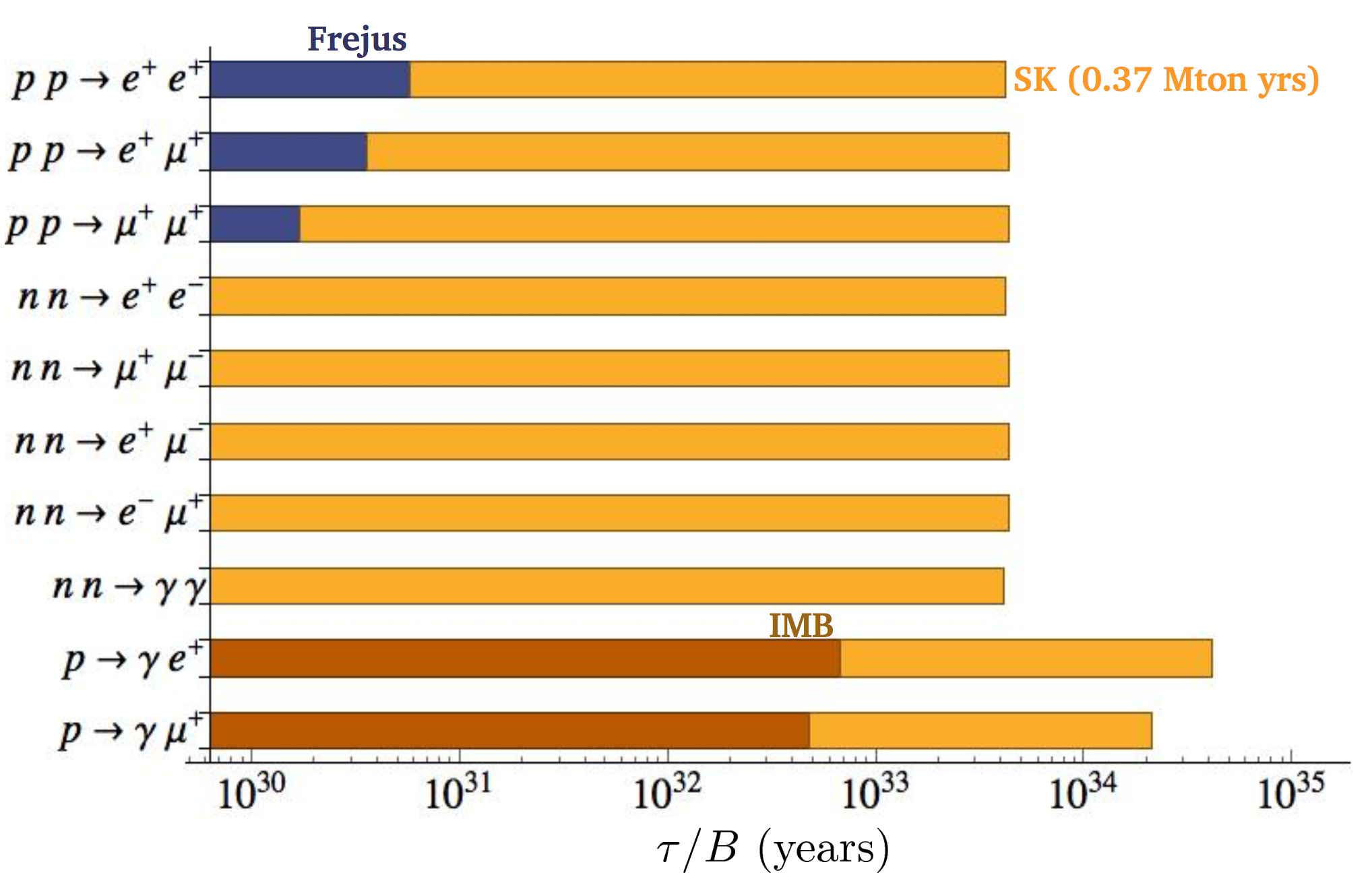}
\caption{(color online) The partial lifetime limits set by Super-Kamiokande for these ten modes, compared with previous limits set by the Fr\'ejus and IMB detectors \cite{frejus, IMB}. Note that Fr\'ejus set dinucleon decay lifetime limits per iron nucleus rather than per oxygen nucleus.} 
\label{fig:moneyplot}
\end{figure}

Therefore we obtain the lower bound on the partial lifetime limit of a decay mode: $\tau/\text{B} = 1/\Gamma_{\text{limit}}$. Table~\ref{tab:lifetimes} summarizes the partial lifetime limits obtained for the ten decay modes studied, and these are also shown in relation to previous measurements in Fig.~\ref{fig:moneyplot}. 

\begin{table}[htbp]
\centering
\begin{tabular}{l @{\hspace{5mm}}c @{\hspace{5mm}}c}
\hline \hline
\noalign{\vskip 1mm}
\multirow{3}{*}{Decay mode} & \multicolumn{2}{c}{Lifetime limit} \\
 & per oxygen nucleus & per nucleon\\
 & ($\times 10^{33}$ years) & ($\times 10^{34}$ years)\\
 \noalign{\vskip 1mm}
 \hline
 \noalign{\vskip 1mm}
 $pp \rightarrow e^+ e^+$ & 4.2 & ---\\
 $nn \rightarrow e^+ e^-$ & 4.2 & --- \\
 $nn \rightarrow \gamma \gamma$ & 4.1 & ---\\ 
 $pp \rightarrow e^+ \mu^+$ & 4.4 & ---\\ 
 $nn \rightarrow e^+ \mu^-$ & 4.4 & ---\\ 
 $nn \rightarrow e^- \mu^+$ & 4.4 & ---\\
 $pp \rightarrow \mu^+ \mu^+$ & 4.4 & ---\\
 $nn \rightarrow \mu^+ \mu^-$ & 4.4 & ---\\ 
 $p \rightarrow e^+ \gamma$ & --- & 4.1 \\ 
 $p \rightarrow \mu^+ \gamma$ & --- & 2.1 \\ 
 \noalign{\vskip 1mm}
 \hline \hline
\end{tabular}
\caption{\label{tab:lifetimes} Summary of the partial lifetime limits for each of the ten dinucleon and nucleon decay modes, including systematic uncertainties, at 90\% CL.}
\end{table}

We searched for the 10 dinucleon and nucleon decay modes characterized by a two-body final state with no hadrons in the Super-Kamiokande data with an accumulated exposure of 0.37 megaton$\cdot$years. No significant evidence for dinucleon or nucleon decay was observed, and we set lower limits on the partial lifetimes that are above $4 \times 10^{33}$ years for the dinucleon decay modes, $4.1 \times 10^{34}$ years for $p \rightarrow e^+ \gamma$, and $2.1 \times 10^{34}$ years for $p \rightarrow \mu^+ \gamma$. For five of the modes, the limits are novel, and the limits for all 10 modes are the most stringent by over one order of magnitude. 

We gratefully acknowledge the cooperation of the Kamioka Mining and Smelting Company. The Super-Kamiokande experiment has been built and operated from funding by the Japanese Ministry of Education, Culture, Sports, Science and Technology, the U.S. Department of Energy, and the U.S. National Science Foundation.

\end{document}